\documentstyle[12pt,epsfig,dina4p]{article}
\newcommand{\ptmiss}{{\mbox{$\not\hspace{-.55ex}{P}_t$}}}

\def\etjet{E_T^{jet}}
\def\etajet{\eta^{jet}}

\def\etcal{E_{T,cal}^{jet}}
\def\etacal{\eta_{cal}^{jet}}

\def\phical{\varphi_{cal}^{jet}}

\def\etaphi{\eta-\varphi}
\def\etar{-1<\etajet<2}
\def\g2{GeV$^2$}
\def\q2{Q^2}
\def\oalphas3{O(\alpha_S^3)}

\begin{document}

\vspace{1 cm}

\begin{titlepage}

\title{{\bf 
 Measurement of Jet Shapes in High-$Q^2$ \\ Deep Inelastic Scattering at
 HERA}
\author{ ZEUS Collaboration }}
\date{}

\maketitle

\vspace{4 cm}

\begin{abstract}

 The shapes of jets with transverse energies, $\etjet$, up to 45~GeV 
produced in neutral- and charged-current deep inelastic $e^+p$ scattering 
(DIS) at $Q^2 > 100$~GeV$^2$ have been measured with the ZEUS detector at 
HERA. Jets are identified using a cone algorithm in the $\etaphi$ plane 
with a cone radius of one unit. The jets become narrower as $E^{jet}_T$
increases. The jet shapes in neutral- and charged-current DIS are found
to be very similar. The jets in neutral-current DIS are narrower than
those in resolved processes in photoproduction and closer to those in 
direct-photon processes for the same ranges in $\etjet$ and jet
pseudorapidity. The jet shapes in DIS are observed to be similar to those
in $e^+e^-$ interactions and narrower than those in $\bar{p}p$ collisions for 
comparable $\etjet$. Since the jets in $e^+e^-$ interactions and $e^+p$ DIS are
predominantly quark initiated in both cases, the similarity in the jet shapes
indicates that the pattern of QCD radiation within a quark jet is to a large
extent independent of the hard scattering process in these reactions.

\end{abstract}
 
\vspace{-18cm}
{\noindent
 DESY 98-038 \newline
 March 1998}
 
\setcounter{page}{0}
\thispagestyle{empty}
\pagenumbering{Roman}
\def\3{\ss}
\parindent0.cm
\parskip 3mm plus 2mm minus 2mm
 
\newpage

\begin{center}
{\Large  The ZEUS Collaboration}
\end{center}

  J.~Breitweg,
  M.~Derrick,
  D.~Krakauer,
  S.~Magill,
  D.~Mikunas,
  B.~Musgrave,
  J.~Repond,
  R.~Stanek,
  R.L.~Talaga,
  R.~Yoshida,
  H.~Zhang  \\ 
 {\it Argonne National Laboratory, Argonne, IL, USA}~$^{p}$
% ...

\newpage

% ...
$^{\    2}$ now at Univ. of Salerno and INFN Napoli, Italy \\ 
% ...
$^{  21}$ now at ATLAS Collaboration, Univ. of Munich \\
% ...
$^{  31}$ partially supported by the Foundation for German-Russian
Collaboration DFG-RFBR \\ \hspace*{3.5mm} (grant no. 436 RUS 113/248/3 and 
no. 436 RUS 113/248/2)\\
% ...

\newpage

\begin{tabular}[h]{rp{14cm}}

$^{a}$ &  supported by the Natural Sciences and Engineering Research
          Council of Canada (NSERC)  \\
$^{b}$ &  supported by the FCAR of Qu\'ebec, Canada  \\
$^{c}$ &  supported by the German Federal Ministry for Education and
          Science, Research and Technology (BMBF), under contract
          numbers 057BN19P, 057FR19P, 057HH19P, 057HH29P \\
$^{d}$ &  supported by the MINERVA Gesellschaft f\"ur Forschung GmbH,
          the German Israeli Foundation, the U.S.-Israel Binational
          Science Foundation, and by the Israel Ministry of Science \\
$^{e}$ &  supported by the German-Israeli Foundation, the Israel Science
          Foundation, the U.S.-Israel Binational Science Foundation, and by
          the Israel Ministry of Science \\
$^{f}$ &  supported by the Italian National Institute for Nuclear Physics 
          (INFN) \\
$^{g}$ &  supported by the Japanese Ministry of Education, Science and
          Culture (the Monbusho) and its grants for Scientific Research \\
$^{h}$ &  supported by the Korean Ministry of Education and Korea Science
          and Engineering Foundation  \\ 
$^{i}$ &  supported by the Netherlands Foundation for Research on
          Matter (FOM) \\
$^{j}$ &  supported by the Polish State Committee for Scientific
          Research, grant No.~115/E-343/SPUB/P03/002/97, 2P03B10512,
          2P03B10612, 2P03B14212, 2P03B10412 \\ 
$^{k}$ &  supported by the Polish State Committee for Scientific
          Research (grant No. 2P03B08308) and Foundation for
          Polish-German Collaboration  \\
$^{l}$ &  partially supported by the German Federal Ministry for
          Education and Science, Research and Technology (BMBF)  \\ 
$^{m}$ &  supported by the Fund for Fundamental Research of Russian Ministry
          for Science and Edu\-cation and by the German Federal Ministry for
          Education and Science, Research and Technology (BMBF) \\
$^{n}$ &  supported by the Spanish Ministry of Education 
          and Science through funds provided by CICYT \\ 
$^{o}$ &  supported by the Particle Physics and 
          Astronomy Research Council \\
$^{p}$ &  supported by the US Department of Energy \\ 
$^{q}$ &  supported by the US National Science Foundation \\ 

\end{tabular}

\end{titlepage}
 
\newpage
\parindent 5mm
\parskip 0mm
\pagenumbering{arabic}
\setcounter{page}{1}
\normalsize

 
\section{Introduction}

 The internal structure of a jet is expected to depend mainly on the type 
of primary parton, quark or gluon, from which it originated and to a lesser
extent on the particular hard scattering process. For cone jet algorithms 
\cite{cone2,snow} a useful representation of the jet's internal structure is
given by the jet shape \cite{sdellis}. At sufficiently high jet energy, where
fragmentation effects become negligible, the jet shape should be calculable by
perturbative Quantum Chromodynamics (pQCD). pQCD predicts gluon jets to be
broader than quark jets as a consequence of the gluon-gluon coupling strength
being larger than that of the quark-gluon coupling \cite{qgjets}. Measurements
of the jet width in $e^+e^-$ interactions at LEP1 have shown that gluon jets
are indeed broader than quark jets \cite{lepjewi}. The dependence of the
structure of quark and gluon jets on the production process can be
investigated by comparing measurements of the jet shape in different reactions
in which the final-state jets are predominantly quark or gluon initiated.

 Measurements of the integrated jet shape were made in $\bar{p}p$ 
collisions at $\sqrt{s}=1.8$~TeV using charged particles \cite{cdf1} as 
well as both neutral and charged particles \cite{d01}, and a qualitative 
agreement with $\oalphas3$ QCD calculations \cite{sdellis,jetrad} was
found. Measurements of the integrated and differential jet shapes were
made in $e^+e^-$ interactions at LEP1 using both neutral and charged
particles \cite{opal1} and were found to be well described by
leading-logarithm parton-shower Monte Carlo calculations. It was observed
\cite{opal1} that the jets in $e^+e^-$ are significantly narrower than those
in $\bar{p}p$ and most of this difference was ascribed to the different
mixtures of quark and gluon jets in the two production processes.

 Measurements of the integrated jet shape in quasi-real photon proton 
collisions at HERA have recently been presented \cite{zshape97} and were found to 
be well described by leading-logarithm parton-shower Monte Carlo 
calculations except for the inclusive production of jets with high jet
pseudorapidity ($\etajet$) and low jet transverse energy ($\etjet$).
Fixed-order perturbative QCD calculations at the parton level \cite{kramer}
are able to describe the measured jet shapes within the uncertainties on
the matching between the theoretical and experimental jet algorithms.

 At HERA, jet production has been observed in both neutral-
\cite{zeusjets,h1jets} and charged-current \cite{zeuscc94} deep inelastic
$ep$ scattering (DIS) at large $\q2$ (where $\q2$ is the virtuality of the 
exchanged boson). In this paper, measurements of the differential and
integrated jet shapes in neutral- and charged-current DIS at $\q2>100$~\g2\ 
are presented. The data sample used in this analysis 
has been collected with the ZEUS detector in $e^+ p$ interactions at the 
HERA collider. To compare with measurements of the jet shapes in $\bar{p}p$,
$\gamma p$ and $e^+e^-$ collisions, jets are searched for with an iterative
cone algorithm \cite{zshape97} with radius $R=1$ in the
pseudorapidity\footnote{The ZEUS coordinate system is defined as 
right-handed with the $Z$-axis pointing in the proton beam direction, 
hereafter referred to as forward, and the $X$-axis horizontal, pointing 
towards the centre of HERA. The pseudorapidity is defined as 
$\eta=-\ln(\tan\frac{\theta}{2})$, where the polar angle $\theta$ is taken
with respect to the proton beam direction.}
($\eta$) - azimuth ($\varphi$) plane of the laboratory frame. Jets have
been selected with jet transverse (with respect to the proton beam 
direction) energy $\etjet>14$~GeV and jet pseudorapidity in the range $\etar$.
The jet shape has been measured using the ZEUS calorimeter and corrected to
the hadron level. The measurements are presented as functions of $\etjet$ and
$\etajet$. The measured jet shapes are compared to similar measurements in
other reactions and to leading-logarithm parton-shower Monte Carlo
calculations.


\section{Experimental setup}

 During 1995 and 1996 HERA operated with protons of energy $E_p = 820$~GeV 
and positrons of energy $E_e = 27.5$~GeV. The ZEUS detector is described in
detail in \cite{sigtot,status}. The main subdetectors used in the present
analysis are the central tracking system positioned in a 1.43~T solenoidal
magnetic field and the uranium-scintillator sampling calorimeter (CAL).
The tracking system was used to establish an interaction vertex and to
select neutral- and charged-current DIS events. The CAL is hermetic and
consists of 5918~cells each read out by two photomultipliers tubes. Under
test beam conditions the CAL has energy resolutions of 18\%/$\sqrt{E}$ for
electrons and 35\%/$\sqrt{E}$ for hadrons. Energy deposits in the CAL were
used to identify the scattered positron, to find jets and to
measure jet energies. Jet energies are corrected for the energy lost in
inactive material in front of the CAL. This material is typically about one
radiation length. The effects of uranium noise were minimised by
discarding cells in the electromagnetic (EMC) or hadronic (HAC) sections
if they had energy deposits of less than 60~MeV or 110~MeV, respectively.
A three--level trigger was used to select events online
\cite{zeuscc94,status,zeushix}.


\section{Data selection}

 Neutral-current (NC) DIS events have been selected offline from the ZEUS 
1995 data sample, which corresponds to an integrated luminosity of
6.3~pb$^{-1}$, using criteria similar to those reported previously 
\cite{zeushix,zeusf294}. The main steps are briefly discussed here. The 
scattered positron candidate has been identified by using the pattern of 
energy deposits in the CAL \cite{sinistra}. The energy ($E_{e^{\prime}}$) 
and polar angle ($\theta_{e^{\prime}}$) of the positron candidate have 
been determined from the CAL measurements. The $Q^2$ variable has been 
reconstructed by the double-angle method ($Q^2_{DA}$) \cite{dameth}, which 
uses $\theta_{e^{\prime}}$ and an angle that corresponds to the direction of
the scattered quark in quark-parton model type events. This second angle
has been determined from the CAL measurements of the hadronic final
state. The following requirements have been imposed:
\begin{itemize}
 \item A positron candidate of uncorrected energy $E_{e^{\prime}}>10$~GeV.
      This cut ensures a high and well understood positron finding
      efficiency and suppresses background from photoproduction events,
      where the scattered positron escapes down the rear beampipe.
 \item $y_e < 0.95$, where $y_e=1-E_{e^{\prime}}
      (1-\cos{\theta_{e^{\prime}}})/(2 E_e)$. This condition removes events where
      fake positron candidates are found in the forward
      region of the CAL.
 \item The total energy not associated with the positron candidate within 
      a cone of radius 0.7 units in the $\etaphi$ plane around the 
      positron direction must be less than 5~GeV. This condition removes
      photoproduction and DIS events where part of a jet has been falsely
      identified as the scattered positron.
 \item A track is required to match the positron candidate identified in the
      CAL for $\eta_e < 2$, where $\eta_e$ is the pseudorapidity 
      of the positron candidate. This requirement suppresses
      cosmic rays, beam-halo muons, photoproduction and DIS events where
      an electromagnetic shower in the CAL has been falsely identified as
      the scattered positron.
 \item For $\eta_e > 2$ the transverse energy of the positron candidate should
      be larger than 20~GeV. This requirement further reduces the number of
      fake positrons in the forward region of the CAL.
 \item 38~GeV~$<(E-p_Z)<65$~GeV, where $E$ is the total energy as measured
      by the CAL, $E=\sum_i E_i$, and $p_Z$ is the $Z$-component of the
      vector $\vec{p} = \sum_i E_i \vec r_i$; in both cases the sum runs
      over all CAL cells, $E_i$ is the energy of the calorimeter cell $i$
      and $\vec r_i$ is a unit vector along the line joining the
      reconstructed vertex and the geometric centre of the cell $i$. This
      cut removes events with large initial-state radiation and further
      reduces the background from photoproduction.
 \item Events have been removed from the sample if there was a second
      positron candidate with energy above 10~GeV and without
      a track match, and the energy in the CAL after subtracting that of
      the two positron candidates is below 10~GeV. This requirement removes
      elastic Compton scattering events ($ep \rightarrow e\gamma p$).
 \item $\ptmiss/\sqrt{E_t} < 3$~GeV$^{1/2}$ where $\ptmiss$ is the
      missing transverse momentum as measured with the CAL
      ($\ptmiss \equiv \sqrt{p_X^2+p_Y^2}$) and $E_t$ is the
      total transverse energy in the CAL. This cut removes
      cosmic rays and beam-related background.
 \item The vertex position along the beam axis must be in the range
      $-30 < Z < 36$~cm.
 \item $Q^2_{DA}>100$~GeV$^2$.
\end{itemize}

 Charged-current (CC) DIS events have been selected offline from the ZEUS
1995 and 1996 data samples, which correspond to an integrated luminosity
of 14.8~pb$^{-1}$, using criteria similar to those reported in
\cite{zeuscc94}. The $Q^2$ variable has been determined using the method of 
Jacquet-Blondel ($Q^2_{JB}$) \cite{jacblo}, which uses the information from 
the hadronic energy flow of the event. The following conditions have been 
imposed:
\begin{itemize}
 \item $\ptmiss > 11$~GeV. This cut ensures high trigger efficiency.
 \item $\ptmiss/E_t > 0.5$. This cut rejects photoproduction and
      beam-related background.
 \item The vertex position along the beam axis should lie in the range
      $-30 < Z < 36$~cm.
 \item At least one track should point to the vertex. This requirement rejects
      cosmic rays and beam-gas interactions.
 \item The number of tracks not associated to the vertex must be less than 20\%
      of the total number of tracks. This cut further reduces the
      background from beam-gas interactions.
 \item The difference $\Delta \varphi$ between the azimuths of the net 
      transverse momentum as measured by the tracks associated with the 
      vertex and as measured by the CAL has been required to 
      fulfill $|\Delta \varphi |< 1$~rad. This requirement removes overlays
      of cosmic rays on $ep$ interactions.
 \item $P^{tracks}_t/\ptmiss > 0.1$, where $P^{tracks}_t$ is the net 
      transverse momentum of the tracks associated with the vertex (this 
      condition has not been applied if $\ptmiss > 25$~GeV). This cut
      rejects beam-related background, in which $\ptmiss$
      is pointing to small polar angles, and events with additional
      non-$ep$ related energy deposits in the CAL (mainly cosmic rays).
 \item The event has been removed from the sample if there was an isolated
      positron candidate with energy above 10~GeV and a track match. This
      condition removes NC DIS events.
 \item Pattern recognition algorithms based on the topology of the CAL
      energy distribution were applied to reject cosmic rays and
      beam-halo muons.
 \item $Q^2_{corr} > 100$~GeV$^2$, where $Q^2_{corr}$ denotes the
      corrected value of $Q^2_{JB}$ as described in \cite{zeuscc94}. The
      resolution in the reconstruction of $Q^2$ is $\approx 25$\%.
\end{itemize}

 A search for jet structure using the CAL cells (see Section~5) has been
performed on both samples (NC and CC DIS), and events with at least one 
jet of `corrected' transverse energy (see Section~5) $\etjet>14$~GeV and
$-1<\etajet <2$ have been retained. The selected sample of NC (CC) DIS
consists of 6926 (231) events containing 7092 (233) jets. In both cases,
the background from photoproduction has been estimated using Monte Carlo
techniques and was found to be below 1\%.


\section{Monte Carlo simulation}

 The response of the detector to jets and the correction factors for the 
jet shapes have been determined from samples of Monte Carlo (MC) events.

 NC and CC DIS events have been generated using the {\sc lepto} program 
\cite{lepto} interfaced to {\sc heracles} \cite{heracles} via {\sc django} 
\cite{django}. The {\sc heracles} program includes photon and $Z^{o}$ 
exchanges and first-order electroweak radiative corrections. The CTEQ4D
\cite{cteq4} NLO proton parton densities have been used. The hadronic
final state is simulated using the colour-dipole model \cite{cdm} including
the leading-order (LO) QCD diagrams as implemented in {\sc ariadne} 
\cite{ariadne} for the QCD cascade. As an alternative, samples of events 
have been generated using the model of {\sc lepto} based on first-order 
QCD matrix elements plus parton-shower ({\sc meps}). For the generation of
the samples with {\sc meps}, the soft colour interactions option has been
switched off.

 In addition, a sample of NC DIS events has been generated using the
{\sc pythia} program \cite{pythia}: a lowest-order electroweak calculation
including initial- and final-state QCD radiation in the leading-logarithm
parton-shower approximation. In this case, events have been generated
using the MRSA \cite{mrsa} set of proton parton densities and the
first-order QCD matrix elements have not been included. In all cases, the
{\sc lund} string model \cite{lund} as implemented in {\sc jetset}
\cite{pythia} is used for modelling the fragmentation into hadrons.

 All MC generated events have been passed through the ZEUS detector and
trigger simulation programs \cite{status}. They have been reconstructed and
analysed by the same program chain as the data.


\section{Jet search and energy corrections}

 An iterative cone algorithm in the $\etaphi$ plane \cite{cone2,snow} 
is used to reconstruct jets from the energy measured in the CAL 
cells for both data and MC generated events, and also from the final-state
hadrons for MC generated events. A detailed description of the algorithm
can be found in \cite{zshape97}. The jets reconstructed from the CAL cell 
energies are called $cal$ jets and the variables associated with them
are denoted by $\etcal$, $\etacal$ and $\phical$. The axis of the jet is
defined according to the Snowmass convention \cite{snow}, where $\etacal$
($\phical$) is the transverse--energy weighted mean pseudorapidity
(azimuth) of all the CAL cells belonging to that jet. The energy sharing
of overlapping jets is dealt with using the following procedure. Two jets
are merged if the overlapping energy exceeds 75\% of the total energy of
the jet with the lower energy; otherwise two different jets are formed
and the common cells are assigned to the nearest jet. The cone radius $R$
used in the jet search is set equal to $1$.

 For the MC generated events, the same jet algorithm is also applied to 
the final-state particles. The jets found are called $hadron$ jets and 
the variables associated with them are denoted by $E^{jet}_{T,had}$, 
$\eta^{jet}_{had}$, and $\varphi^{jet}_{had}$. $Hadron$ jets with
$E^{jet}_{T,had}>14$~GeV and $-1<\eta^{jet}_{had}<2$ are selected.

 The comparison of the reconstructed jet variables between the $hadron$ and
the $cal$ jets in MC generated events \cite{zeoct94} shows no significant 
systematic shift in the angular variables $\etacal$ and $\phical$ with 
respect to $\eta^{jet}_{had}$ and $\varphi^{jet}_{had}$. However, the
transverse energy of the $cal$ jet underestimates that of the $hadron$ jet
by an average amount of $\approx$~16\% with an r.m.s. of 11\%. This effect
is due mainly to energy losses in the inactive material in front of the
CAL and is corrected for using the following procedure. The transverse 
energy corrections to $cal$ jets averaged over the azimuthal angle are 
determined using the samples of MC generated events \cite{zeoct94}. These 
corrections are constructed as multiplicative factors, $C(\etcal,\etacal)$, 
which, when applied to the $E_T$ of the $cal$ jets, give the `corrected' 
transverse energies of the jets, $\etjet=C(\etcal,\etacal) \times \etcal$ 
\cite{zeoct94}.


\section{Jet shape}

 The differential jet shape is defined as the average fraction of the
jet's transverse energy that lies inside an annulus in the $\etaphi$ plane
of inner (outer) radius $r-\Delta r/2$ ($r+\Delta r/2$) concentric with the jet
defining cone \cite{sdellis}:
\begin{equation}
\label{eqjsd}
 \rho(r)=\frac{1}{N_{jets}} \frac{1}{\Delta r}
 \sum_{jets}\frac{E_T(r-\Delta r/2,r+\Delta r/2)}{E_T(0,R)},
\end{equation}
where $E_T(r-\Delta r/2,r+\Delta r/2)$ is the transverse energy within the 
given annulus and $N_{jets}$ is the total number of jets in the sample.
The differential jet shape has been measured for $r$ values varying from 
$0.05$ to $0.95$ in $\Delta r=0.1$ increments. The integrated jet shape 
defined by
\begin{equation}
\label{eqjsi}
  \psi(r) = \frac{1}{N_{jets}} \sum_{jets} \frac{E_T(0,r)}{E_T(0,R)}
\end{equation}
is also used. By definition, $\psi(R)=1$. It has been measured for $r$
values varying from $0.1$ to $1.0$ in $\Delta r=0.1$ increments.

 The following procedure is used to reconstruct the differential jet
shape from the CAL cells in data and MC generated events: for each jet
the sum of the transverse energies of the CAL cells assigned to the jet, 
$E_{T,cal}(r-\Delta r/2,r+\Delta r/2)$, with a distance 
$r'=\sqrt{(\Delta \eta)^2+(\Delta \varphi)^2}$ to the jet axis between
$r-\Delta r/2$ and $r+\Delta r/2$ is determined and divided by
$E_{T,cal}(0,1)$. The differential jet shape as measured with the CAL,
$\rho_{cal}(r)$, is then defined in analogy with eq.~(\ref{eqjsd}), where
the sum now runs over all the $cal$ jets in the selected sample and 
$N_{jets}$ is the total number of $cal$ jets in the sample. Similarly, 
the integrated jet shape as measured with the CAL, $\psi_{cal}(r)$, is
defined in analogy with eq.~(\ref{eqjsi}).

 The same jet shape definition as used above for the CAL cells is applied
to the final-state particles in the case of MC generated events and the
resulting differential (integrated) jet shape is denoted by
$\rho^{MC}_{had}(r)$ ($\psi^{MC}_{had}(r)$).

\subsection{Jet shape correction}

 The differential and integrated jet shapes as measured with the CAL are
corrected back to the hadron level using the samples of MC generated
events. The corrected differential and integrated jet shapes, $\rho(r)$ and
$\psi(r)$, refer to jets at the hadron level with a cone radius of one
unit in the $\etaphi$ plane. The measurements are given for jets with 
$\etjet > 14$~GeV and $ -1 < \etajet < 2 $ in the kinematic region $Q^2 > 100$ GeV$^2$. 

 The corrected jet transverse energy is used only to select the sample
of jets ($\etjet > 14$~GeV) and to study the dependence of the jet shape
as a function of $\etjet$. The reconstructed jet shapes are then corrected
for acceptance and smearing effects using the samples of MC generated
events. The correction factors also take into account the efficiency of
the trigger, the selection criteria, the purity and efficiency of the jet
reconstruction, and the effects of the energy losses due to inactive
material in front of the CAL.
The corrected differential (integrated) jet shape is determined
bin-by-bin as $\rho(r) = G^{MC}_{cal}(r) \cdot \rho_{cal}(r)$ and
$\psi (r) = F^{MC}_{cal}(r) \cdot \psi_{cal}(r)$, where the correction 
factors are defined as $G^{MC}_{cal}(r) = \rho^{MC}_{had}(r)/\rho^{MC}_{cal}(r)$ and
$F^{MC}_{cal}(r) = \psi^{MC}_{had}(r)/\psi^{MC}_{cal}(r)$
and are determined separately for each region of $\etajet$ and $\etjet$. 

 For this approach to be valid, the uncorrected jet shapes in the data 
must be described by the MC simulations at the detector level. 
As shown later, this condition is satisfied by the {\sc ariadne} and 
{\sc meps} simulations in all $\etajet$ and $\etjet$ regions studied. 
The samples of events generated with {\sc ariadne} are used to correct 
the jet shapes. The correction factors $G^{MC}_{cal}(r)$ do not show a 
strong dependence on $\etajet$ or $\etjet$ and vary between $0.7$ and $1$
for $r \geq 0.15$. The correction factors for the integrated jet shape
$F^{MC}_{cal}(r)$ differ from unity by less than 25\% for $r\geq0.2$.
Close to the centre of the jet the correction factor $G^{MC}_{cal}(r=0.05)
\equiv F^{MC}_{cal}(r=0.1)$ is large and varies between $1.4$ and $1.7$
depending on $\etajet$ and $\etjet$.

 The jet shapes have been also reconstructed using tracks instead of CAL
cells both in data and MC generated events. Since the use of tracks gives
an improved spatial resolution for the transverse-energy flow of the
charged particles within a jet, this study provides a cross-check of the
resolution in $r$ for the jet shape reconstructed using the CAL. The
resulting corrected jet shapes are consistent with those using the CAL
cells within the uncertainties of the measurements (see next section).

\subsection{Systematic uncertainties}

 A detailed study of the sources contributing to the systematic
uncertainties of the measurements has been carried out \cite{mmthesis}. The
uncertainties have been classified into four groups: 
\begin{itemize}
 \item The energy corrections to the jets and the correction functions to the
       jet shapes in NC and CC DIS have been evaluated using the {\sc meps}
       generator. The changes induced in $\rho(r)$ are typically below 10\%.
 \item The absolute energy scale of the $cal$ jets in the MC generated events 
       has been varied by $\pm 3$\%. The resulting corrected $\rho(r)$
       changes typically by less than 3\%.
 \item Variations in the simulation of the CAL response to low-energy 
       particles yielded changes in $\rho(r)$ typically below 3\%.
 \item Variations in the simulation of the trigger and a variation of the
       cuts used to select the data within the ranges allowed by the 
       comparison between data and MC simulations resulted in 
       negligible changes in the corrected jet shapes.
\end{itemize}

 For the measurements of jet shapes in NC DIS, the statistical errors are 
negligible compared to the systematic uncertainties. Conversely, the
statistical errors dominate in the CC DIS analysis. The total positive 
(negative) systematic uncertainty on $\rho(r)$ at each value of $r$ has 
been determined by adding in quadrature the positive (negative) deviations
from the central value. The systematic uncertainties have been added
in quadrature to the statistical errors and are shown as error bars in 
the figures.


\section{Results}

\subsection{Jet shapes in DIS}

 The differential and integrated jet shapes are measured for jets 
in the reactions
$$ e^+ p \rightarrow e^+ (\overline{\nu}) \; + \; {\rm jet} \; + \; X $$
with $Q^2>100$~GeV$^2$. Jets are required to have $\etjet > 14$~GeV and
$-1 < \etajet < 2$. There are 6018, 855 and 53 events in the NC DIS data
sample with $Q^2$ in the range $100-1000$~GeV$^2$, $1000-5000$~GeV$^2$
and $5000-25000$~GeV$^2$. The corresponding numbers for the CC DIS data
sample are 84, 123 and 24 events.

\vspace{0.5cm}
\noindent {\bf Jet shapes in NC DIS}

\begin{figure}[p]
\centerline{
\psfig{figure=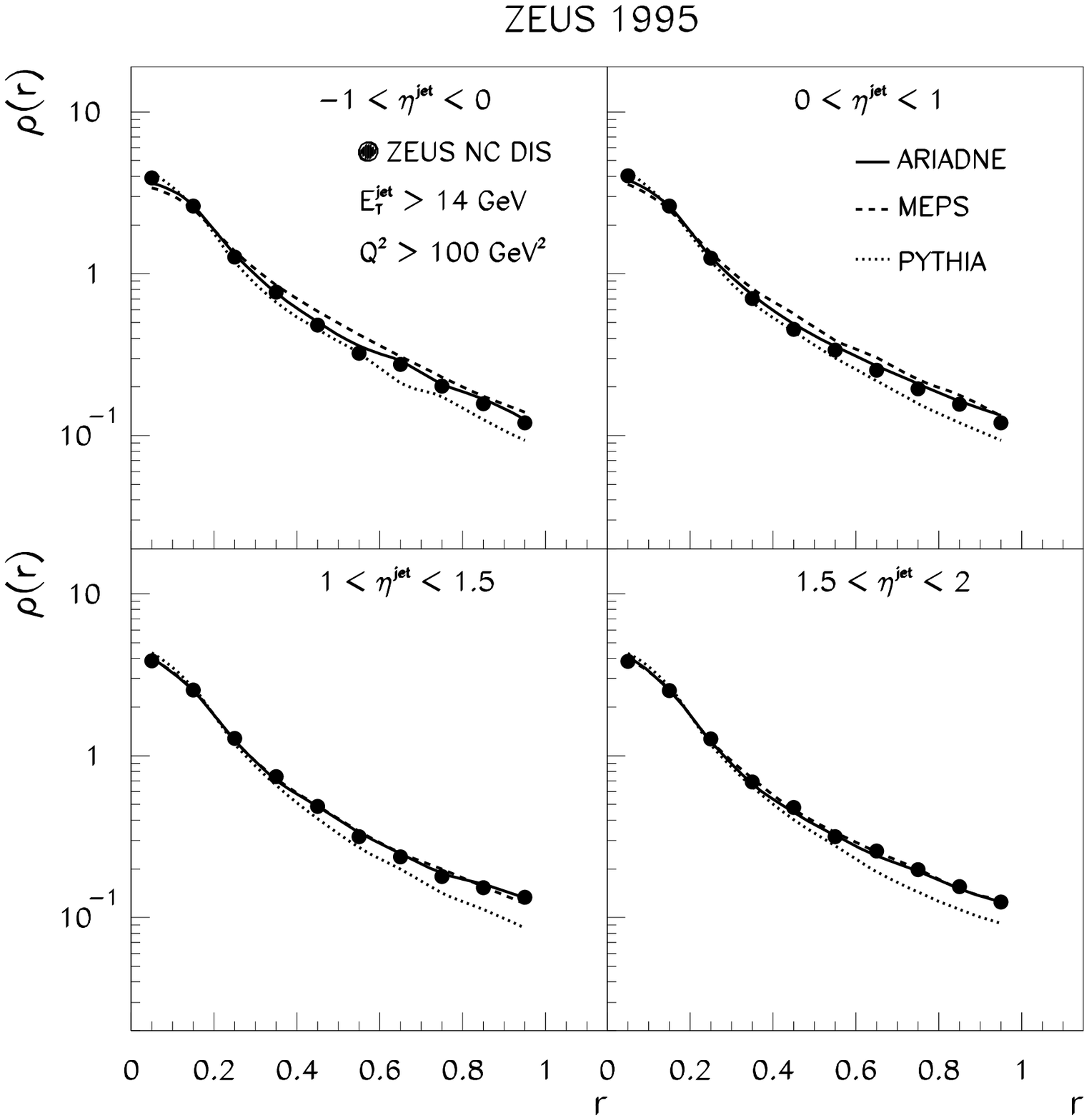,height=15cm}} 
\caption{Measured differential jet shapes corrected to the hadron level, 
 $\rho(r)$, in neutral-current DIS with $\q2 >100$~\g2\ for jets with 
 $E^{jet}_T$ above 14~GeV in different $\eta^{jet}$ regions (black dots).
 The error bars include the statistical and systematic uncertainties 
 added in quadrature (typically smaller than the dots). The predictions of
 {\sc pythia} (dotted lines), {\sc ariadne} (solid lines), and 
 {\sc meps} (dashed lines) are shown for comparison. The predictions have
 been obtained by an integration over the same bins as for the data
 and are presented as smooth curves joining the calculated points.
\label{figdra1}}
\end{figure}

\begin{figure}[p]
\centerline{
\psfig{figure=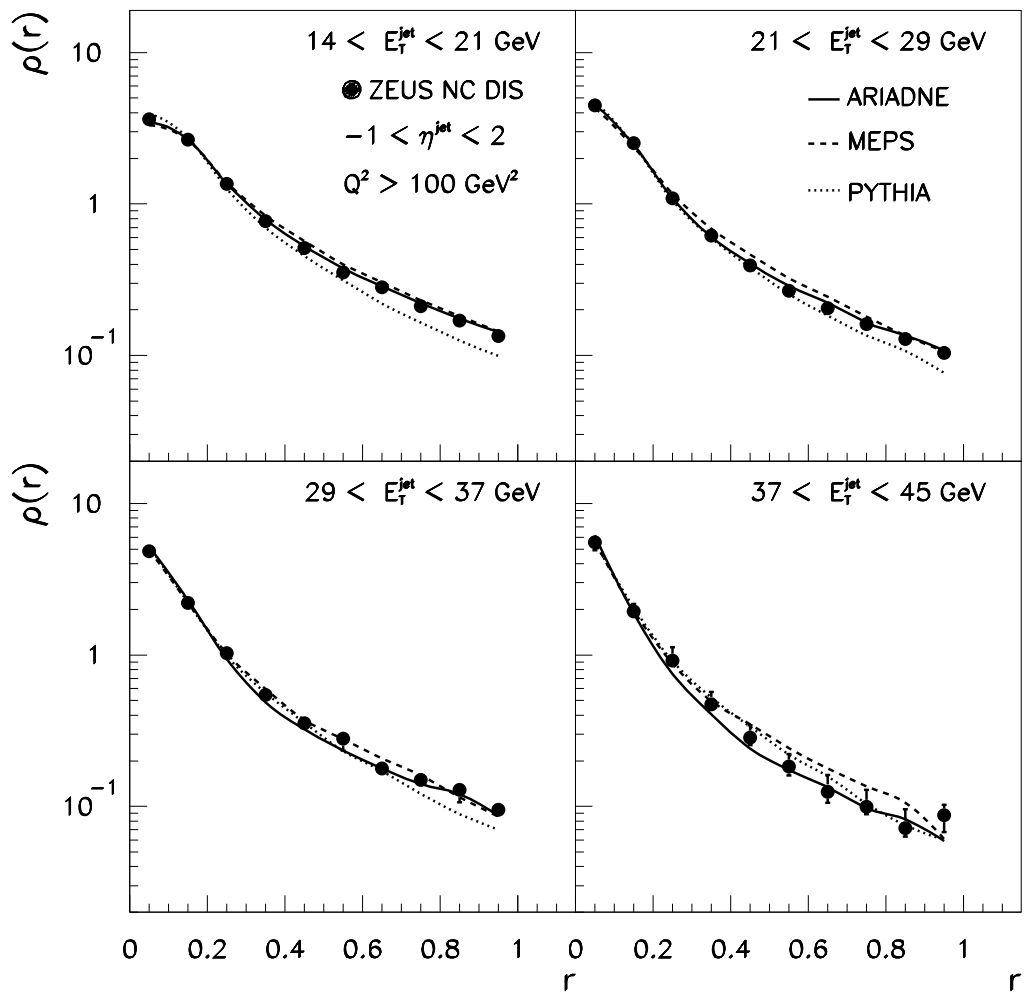,height=15cm}}
\caption{Measured differential jet shapes corrected to the hadron level, 
 $\rho(r)$, in neutral-current DIS with $\q2 > 100$~\g2\ for jets in the 
 $\etajet$ range between $-1$ and 2 in different $\etjet$ regions (black 
 dots). The error bars include the statistical and systematic errors 
 added in quadrature. The predictions of {\sc pythia} (dotted lines), 
 {\sc ariadne} (solid lines), and {\sc meps} (dashed lines) are
 shown for comparison. The predictions have
 been obtained by an integration over the same bins as for the data
 and are presented as smooth curves joining the calculated points.
\label{figdra2}}
\end{figure}

\begin{figure}[p]
\centerline{\psfig{figure=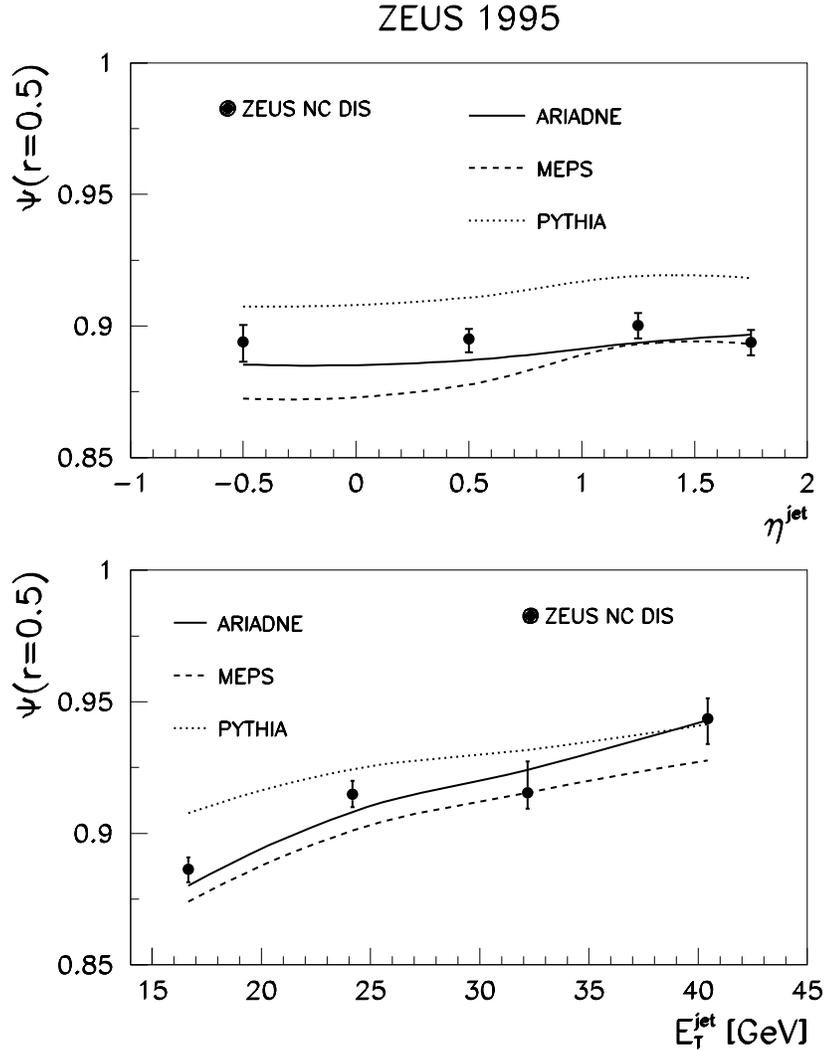,height=15cm}} 
\caption{The measured integrated jet shape corrected to the hadron level
 at a fixed value of $r=0.5$, $\psi(r=0.5)$, as a function of $\etajet$
 (upper plot) and $\etjet$ (lower plot), in neutral-current DIS with
 $\q2 > 100$~\g2\ for jets with $\etjet > 14$~GeV in the $\etajet$ range 
 between $-1$ and 2 (black dots). The error bars include the statistical 
 and systematic errors added in quadrature. The predictions of
 {\sc pythia} (dotted lines), {\sc ariadne} (solid lines), and {\sc meps}
 (dashed lines) are shown for comparison. The predictions have
 been obtained by an integration over the same bins as for the data
 and are presented as smooth curves joining the calculated points.
\label{figdra3}}
\end{figure}

 The measured differential jet shapes in NC DIS for different regions in 
$\etajet$ and $\etjet$ are shown in Figures~\ref{figdra1} and
\ref{figdra2}, respectively. The differential jet shape exhibits a prominent
peak at the centre of the jet. It decreases by a factor $\approx 40$
from the centre of the jet ($r=0.05$) to the edge of the jet ($r=0.95$).
Figure~\ref{figdra3} shows the measured average fraction of the jet's
transverse energy that lies inside an inner cone of radius $r=0.5$
concentric with the jet defining cone, $\psi(r=0.5)$, as functions of
$\etajet$ and $\etjet$. Note that $\psi(r=0.5)$ has been measured in ranges of
$\etjet$ and the data points in Figure~\ref{figdra3} (lower plot) are located
at the weighted mean in each $\etjet$ range. It is observed that the jets
become narrower as $\etjet$ increases. The measured $\psi(r=0.5)$ exhibits no
significant dependence on $\etajet$. 

 The predictions of {\sc ariadne}, {\sc meps} and {\sc pythia} are 
compared to the measured jet shapes in Figures~\ref{figdra1}
to \ref{figdra3}. The predicted jet shape of the colour-dipole model 
({\sc ariadne}) describes the measured jet shape well in all $\etajet$ and
$\etjet$ regions considered. The predicted jets of {\sc pythia} tend to be
narrower at low $\etjet$ than those in the data (see Figure~\ref{figdra3}). 
In the case of {\sc meps}, the predicted jets show a tendency to be
broader at low $\etajet$ than those in the data.

\vspace{0.5cm}
\noindent {\bf Jet shapes in CC DIS}

\begin{figure}[p]
\centerline{
\psfig{figure=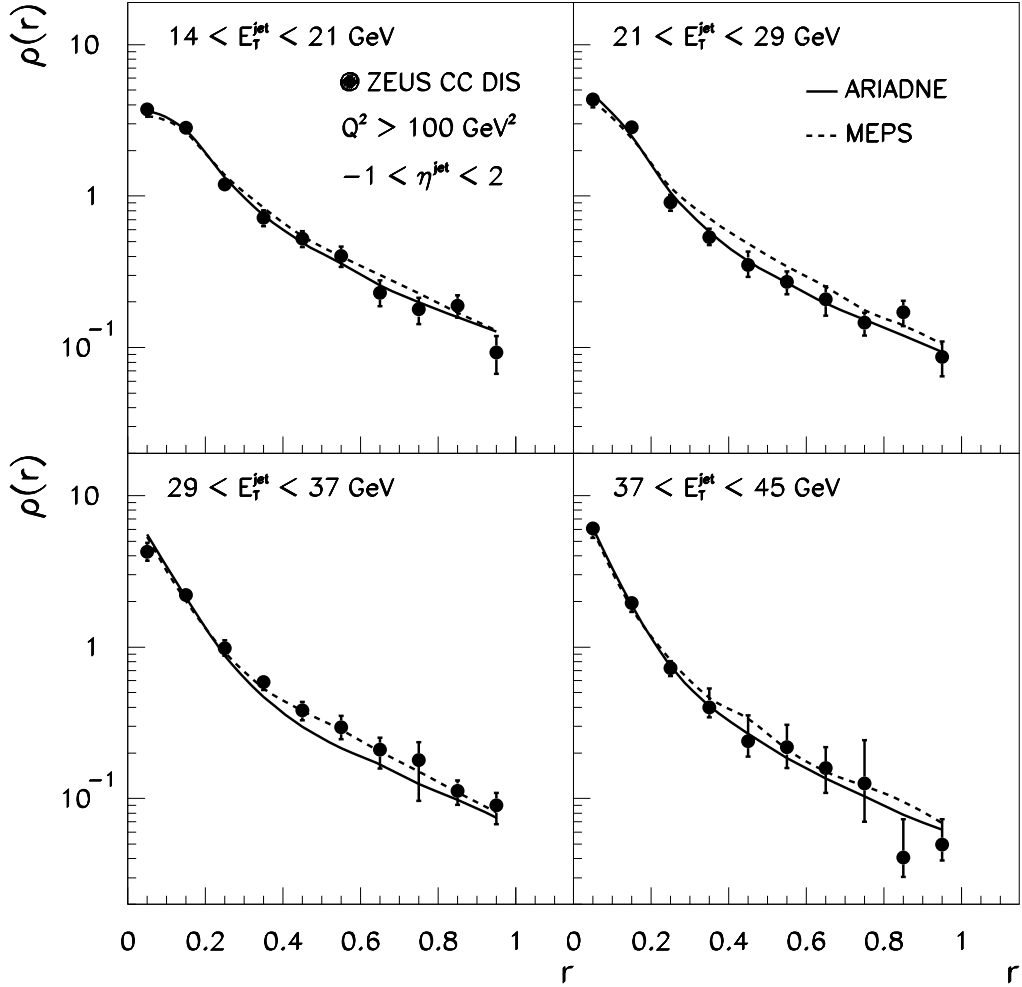,height=15cm}}
\caption{Measured differential jet shapes corrected to the hadron 
 level, $\rho(r)$, in charged-current DIS with $\q2 > 100$~\g2\ for jets
 with $-1<\eta^{jet}<2$ in different $E_T^{jet}$ regions (black dots). 
 The predictions of {\sc ariadne} (solid lines) and {\sc meps}
 (dashed lines) are shown for comparison. The predictions have
 been obtained by an integration over the same bins as for the data
 and are presented as smooth curves joining the calculated points.
\label{figdra4}}
\end{figure}

\begin{figure}[p]
\centerline{
\psfig{figure=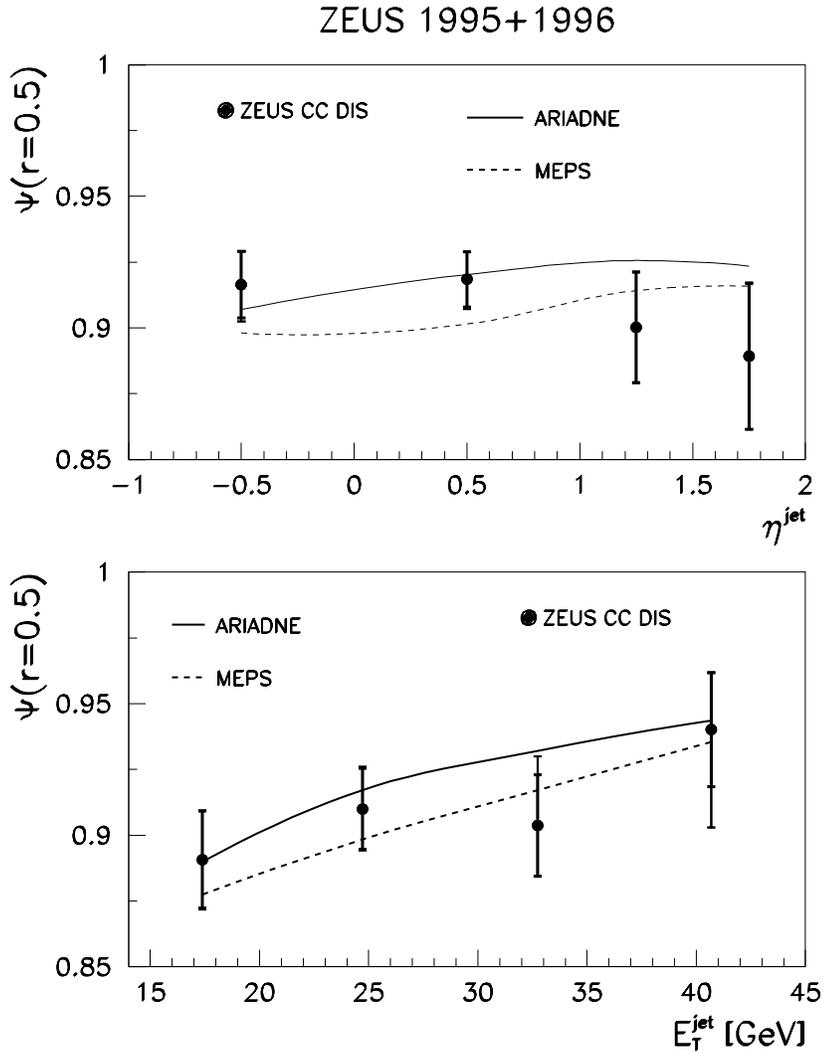,height=15cm}} 
\caption{The measured integrated jet shape corrected to the hadron level
 at a fixed value of $r=0.5$, $\psi(r=0.5)$, as a function of $\etajet$
 (upper plot) and $\etjet$ (lower plot), in charged-current DIS with
 $\q2 > 100$~\g2\ for jets with $\etjet > 14$~GeV in the $\etajet$ range 
 between $-1$ and 2 (black dots). The inner error bars represent the
 statistical errors of the data, and the outer errors bars show the
 statistical and systematic uncertainties added in quadrature. The
 predictions of {\sc ariadne} (solid lines) and {\sc meps} (dashed lines)
 are shown for comparison. The predictions have
 been obtained by an integration over the same bins as for the data
 and are presented as smooth curves joining the calculated points.
\label{figdra5}}
\end{figure}

 The results for $\rho(r)$ in CC DIS for different regions of $\etjet$ are
shown in Figure~\ref{figdra4}. The differential jet shape shows similar
general features to those of the jets in NC DIS.
Figure~\ref{figdra5} shows the measured $\psi(r=0.5)$ as functions of
$\etajet$ and $\etjet$. The measured $\psi(r=0.5)$ exhibits no
significant dependence on $\etajet$ and its dependence on $\etjet$ is
similar to that observed in NC DIS. The predictions of {\sc ariadne} and 
{\sc meps} (see Figures~\ref{figdra4} and \ref{figdra5}) provide a
reasonable description of the measured jet shape.

\begin{figure}[p]
\centerline{
\psfig{figure=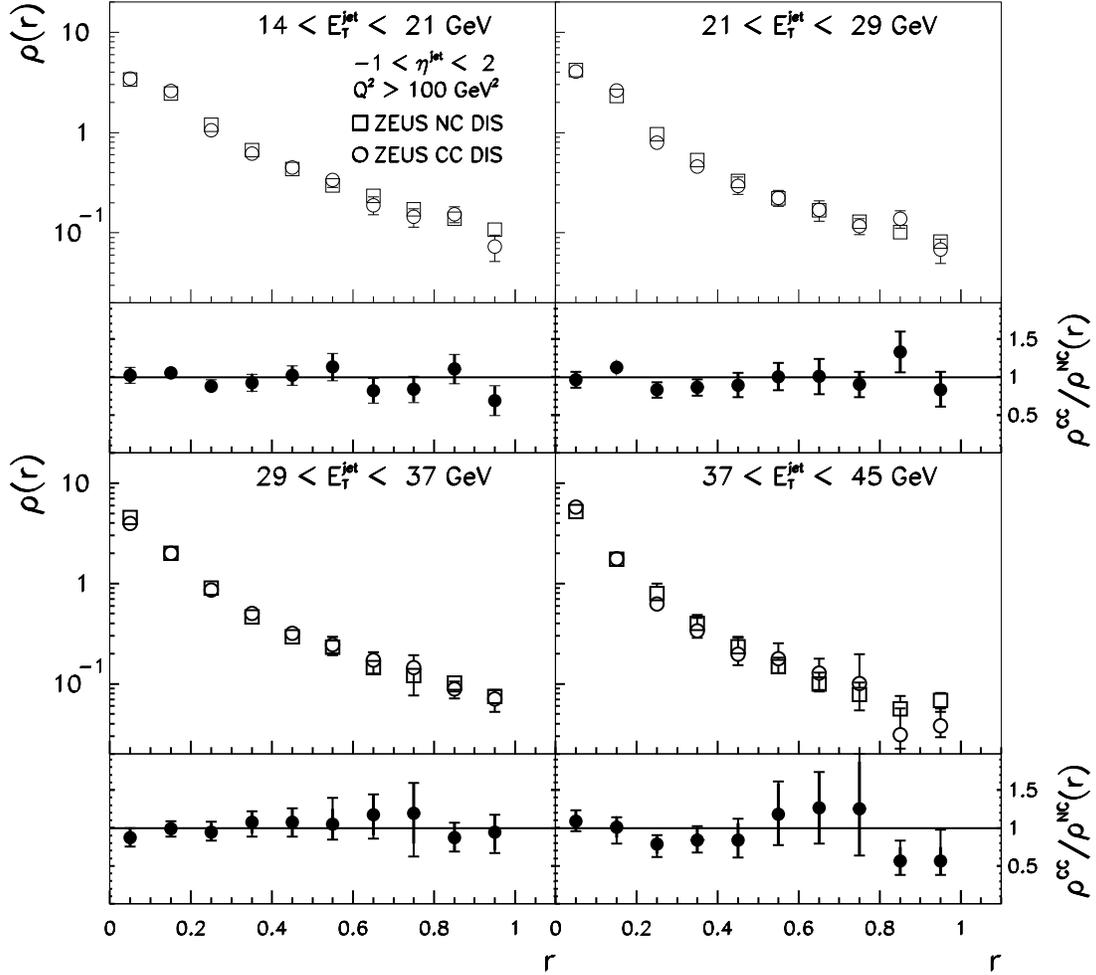,height=15cm}}
\caption{Measured differential jet shapes corrected to the hadron 
 level, $\rho(r)$, in charged-current DIS with $\q2 > 100$~\g2\ for jets
 with $-1<\eta^{jet}<2$ in different $E_T^{jet}$ regions (open circles). 
 The measured jet shapes corrected to the hadron level for jets in
 neutral-current DIS with $\q2 > 100$~\g2\  with $-1<\eta^{jet}<2$ are shown
 for comparison (open squares). Measurements of the ratio
 $\rho^{CC}(r)/\rho^{NC}(r)$ are shown underneath each plot. The inner
 error bars represent the statistical errors of the data, and the outer
 errors bars show the statistical and systematic uncertainties added in
 quadrature.
\label{figdra6}}
\end{figure}

 The measured jet shapes in NC DIS are compared with those in 
CC DIS in Figure~\ref{figdra6} and found to be very similar in each region 
of $\etjet$. Measurements of the ratio of the differential jet shapes in CC
and NC DIS, $\rho^{CC}(r)/\rho^{NC}(r)$, for the same regions of $\etjet$
as above are also shown in Figure~\ref{figdra6} (lower part of each
plot) and found to be compatible with unity. In these measurements some of
the systematic uncertainties common to NC and CC DIS cancel. The
median of the $Q^2$ distribution has been determined for the NC and CC DIS
samples of jets in each $\etjet$ region:
 310~GeV$^2$  (450~GeV$^2$) for $14<\etjet<21$~GeV,
 710~GeV$^2$ (1000~GeV$^2$) for $21<\etjet<29$~GeV, 
1260~GeV$^2$ (1600~GeV$^2$) for $29<\etjet<37$~GeV and
2000~GeV$^2$ (2200~GeV$^2$) for $37<\etjet<45$~GeV in the NC (CC) DIS samples
of jets. Some differences are observed in the $Q^2$ distributions of the
two processes for a given range in $\etjet$. As a cross-check, the jet
shapes in NC and CC DIS have been measured in a common region of $Q^2$
for each range in $\etjet$ and no significant difference has been found.
Therefore, the observation that the jet shapes in NC and CC DIS are very
similar, for the same range of $\etjet$, is independent of the different
$Q^2$ distributions in these processes.

\subsection{Comparison to jet shapes in photoproduction}

 In photoproduction, two types of QCD processes contribute to jet 
production at LO \cite{owens,drees}: either the photon interacts directly 
with a parton in the proton (direct process) or the photon acts as a 
source of partons which interact with those in the proton (resolved 
process). It has been noted that resolved processes dominate jet 
photoproduction in the entire $\etjet$ region studied \cite{zeoct94}. In the 
case of dijet photoproduction the contributions of resolved and direct 
processes can be separated \cite{zenov93} by using the variable
$x^{OBS}_{\gamma}=(\sum_{jets}E^{jet}_T e^{-\eta^{jet}})
/(2 E_{\gamma})$, where the sum runs over the two jets of highest
$E^{jet}_T$ and $E_{\gamma}$ is the initial photon energy. This variable
represents the fraction of the photon's momentum participating in the
production of the two jets with highest $E^{jet}_T$. The LO direct and
resolved processes largely populate different regions of $x^{OBS}_{\gamma}$,
with the direct processes being concentrated at high values. 

\begin{figure}[p]
\centerline{\psfig{figure=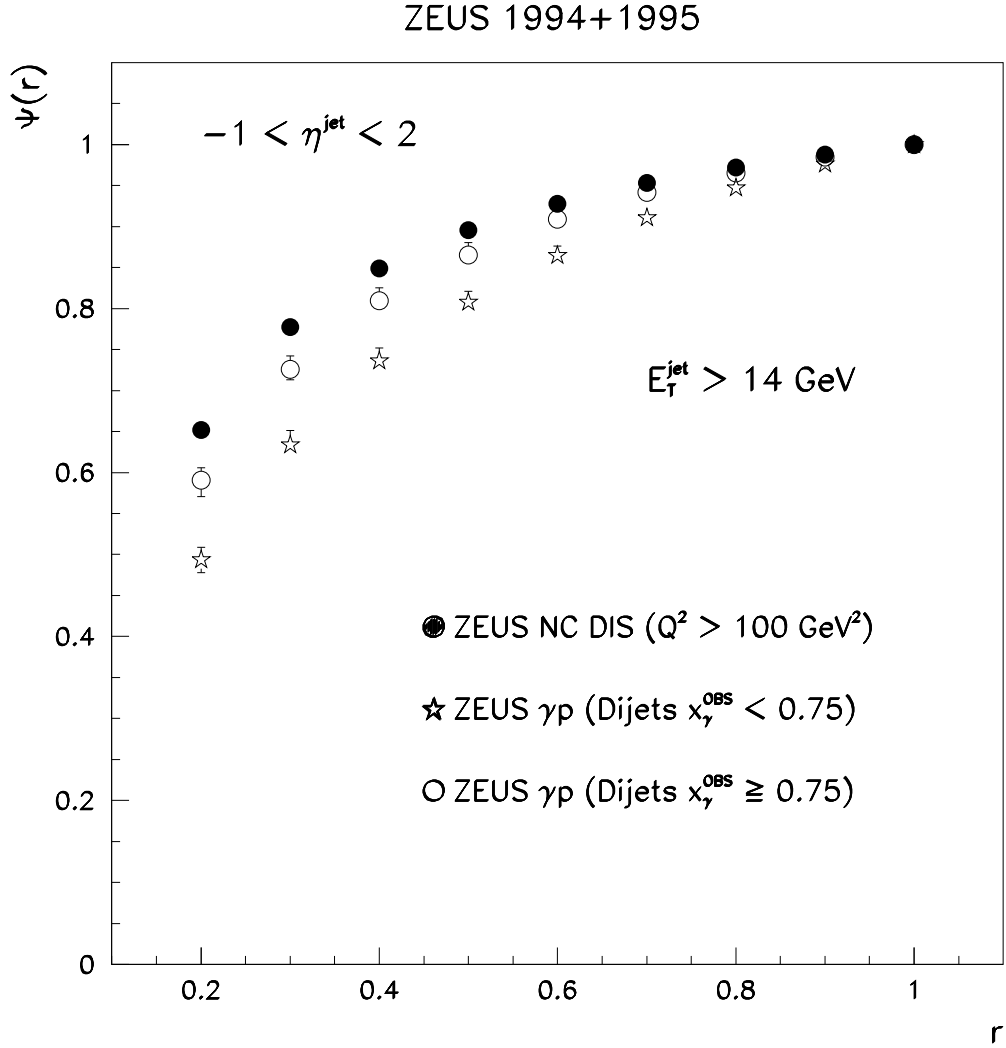,height=15cm}}
\caption{Measured integrated jet shape corrected to the hadron level, 
 $\psi(r)$, in neutral-current DIS with $\q2 > 100$~\g2\ for jets with 
 $\etjet$ above 14~GeV and $-1<\etajet<2$ (black dots). The measured jet
 shape corrected to the hadron level for jets in dijet photoproduction
 with $\etjet$ above 14~GeV and $-1<\etajet<2$ is shown for comparison:
 for dijet production with $x^{OBS}_{\gamma} < 0.75$ (stars) and for
 dijet production with $x^{OBS}_{\gamma}\geq 0.75$ (open circles).
\label{figdra7}}
\end{figure}

 In Figure~\ref{figdra7} the measured integrated jet shape in NC DIS are
compared to those in dijet photoproduction \cite{zshape97} for two different
regions: $x^{OBS}_{\gamma}\geq 0.75$ and $x^{OBS}_{\gamma}< 0.75$. The comparison 
between the jet shapes in NC DIS and dijet photoproduction is made for the
same ranges of $\etajet$ and the $\etjet$ spectrum is similar in these two
processes. The jets produced in NC DIS are narrower than those in dijet
photoproduction but closer to those dominated by direct processes 
($x^{OBS}_{\gamma}\geq 0.75$). This comparison can be understood in terms of
the large fraction of final-state quark jets expected in NC DIS
($e^+ q \rightarrow e^+ q$) and direct processes in photoproduction
(dominated by the subprocess $\gamma g\rightarrow q\bar{q}$). The remaining
differences may be attributed to the contribution from the direct subprocess
$\gamma q\rightarrow qg$ and that of resolved processes, in which the jets are
broader as shown by the measurements in dijet photoproduction with
$x^{OBS}_{\gamma}<0.75$.

\subsection{Comparison to measurements in $e^+e^-$ and $\bar{p}p$
collisions}

 The measured jet shape in NC (CC) DIS with $Q^2 > 100$~GeV$^2$ for jets
with transverse energy between 37 and 45~GeV, with a mean of 40~GeV
(41~GeV), is compared to the measurements of the jet shape corrected to the
hadron level in $\bar{p}p$ collisions by CDF \cite{cdf1} and D\O\ \cite{d01}
and in $e^+e^-$ interactions by OPAL \cite{opal1}:
\begin{itemize}
 \item The CDF data \cite{cdf1} have been obtained using an iterative cone 
      algorithm with $R=1$ similar to that used here.
      The measurements shown are for jets with transverse energy 
      between 40 and 60~GeV, with a mean of 45~GeV, and pseudorapidity
      $0.1<|\etajet|<0.7$. The contribution to the jet shape due to the
      underlying event was found to be small\footnote{In NC and CC DIS the
      underlying event is not expected to contribute in the kinematic
      region studied here.}. If a jet shares more than 75\% of its energy
      with a jet of higher energy, the two are merged together; otherwise,
      they are defined as distinct and the particles common to both jets
      are assigned to the nearest jet.
 \item The D\O\ data \cite{d01} have been obtained also using an iterative
      cone algorithm with $R=1$ similar to that used here.
      The jet direction was defined according to a convention
      different from that of Snowmass; however, this difference is not
      expected to have a significant effect on the results \cite{glover}.
      The jet shape has been measured for jets with transverse energy
      between 45 and 70~GeV, with a mean of 53~GeV, and pseudorapidity
      $|\etajet|<0.2$. The jet shape has been corrected to remove the
      small contribution due to the underlying event. Two jets were
      merged if more than 50\% of the $E_T$ of the jet with smaller $E_T$
      was contained in the overlap region; otherwise, the two jets were
      not merged and each particle in the overlap region was assigned to
      the nearest jet.
 \item The OPAL data \cite{opal1} have been obtained using a cone algorithm
      especially designed to emulate that of the CDF measurements, i.e.
      defining the cone in the $\etaphi$ plane, using $R=1$, demanding 
      $|\etajet|<0.7$ and measuring the transverse energy flow.
      The jet shape has been measured for jets with energy
      greater than 35~GeV, with a mean of 40.4~GeV. The $e^+e^-$ data 
      have no underlying event. Overlapping jets are treated using the
      same procedure as CDF.
\end{itemize}

\begin{figure}[p]
\centerline{\psfig{figure=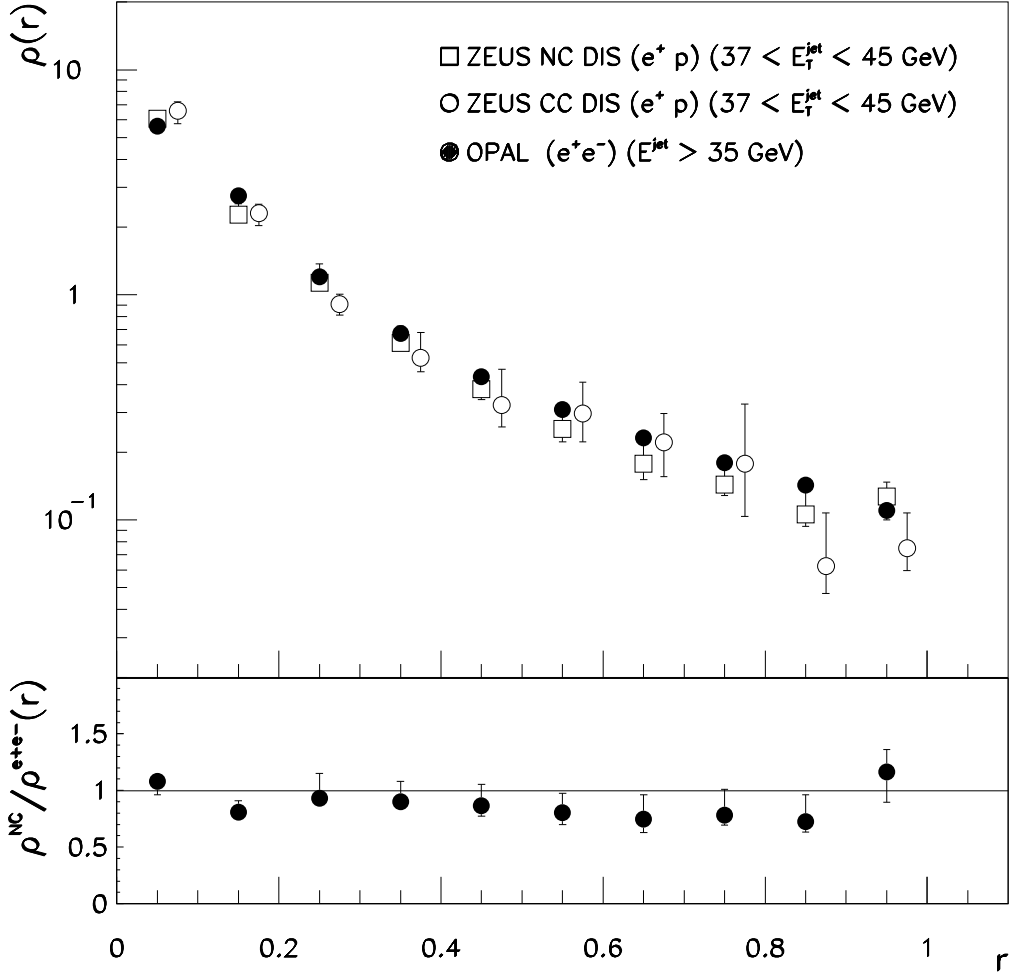,height=15cm}} 
\caption{Measured differential jet shapes corrected to the hadron 
 level, $\rho(r)$, in neutral- (charged-) current DIS with $\q2 > 100$~\g2\
 and a median of $2000$~GeV$^2$ ($2200$~GeV$^2$)
 for jets with $\etajet$ in the range between $-1$ and $2$ and
 $37< \etjet < 45$~GeV are shown as squares (open circles). The
 measurements in CC DIS have been obtained for the same values of $r$ as
 those in NC DIS, and for an easier comparison the measurements are
 plotted at $r+0.025$. The measurements of the jet shape in $e^+e^-$
 interactions by OPAL (black dots) is shown for comparison. The ratio of
 differential jet shapes in NC DIS and $e^+e^-$ interactions,
 $\rho^{NC}(r)/\rho^{e^+e^-}(r)$, is shown in the lower part of the
 figure.
\label{figdra8}}
\end{figure}

 The measured differential jet shapes in NC and CC DIS are compared to that
measured in $e^+e^-$ interactions in Figure~\ref{figdra8} and are found to be
similar. The ratio of the differential jet shapes in NC DIS and $e^+e^-$
interactions, $\rho^{NC}(r)/\rho^{e^+e^-}(r)$, is also shown in Figure~\ref{figdra8}
(lower part of the figure) and is found to be compatible with unity within
the uncertainties of the DIS measurements, which are dominant. For the
selected samples of jets, the jet shapes in $e^+e^-$ interactions and DIS
are expected to be similar due to the large fraction of final-state quark
jets in these two processes. However, some differences may appear since
there are configurations of colour flow (for example, that of
initial-state QCD radiation) in DIS which are not present in $e^+e^-$. The
striking similarity in the jet shapes indicates the large extent to which
the pattern of QCD radiation within a quark jet is independent of
the hard scattering process in these reactions.

\begin{figure}[p]
\centerline{\psfig{figure=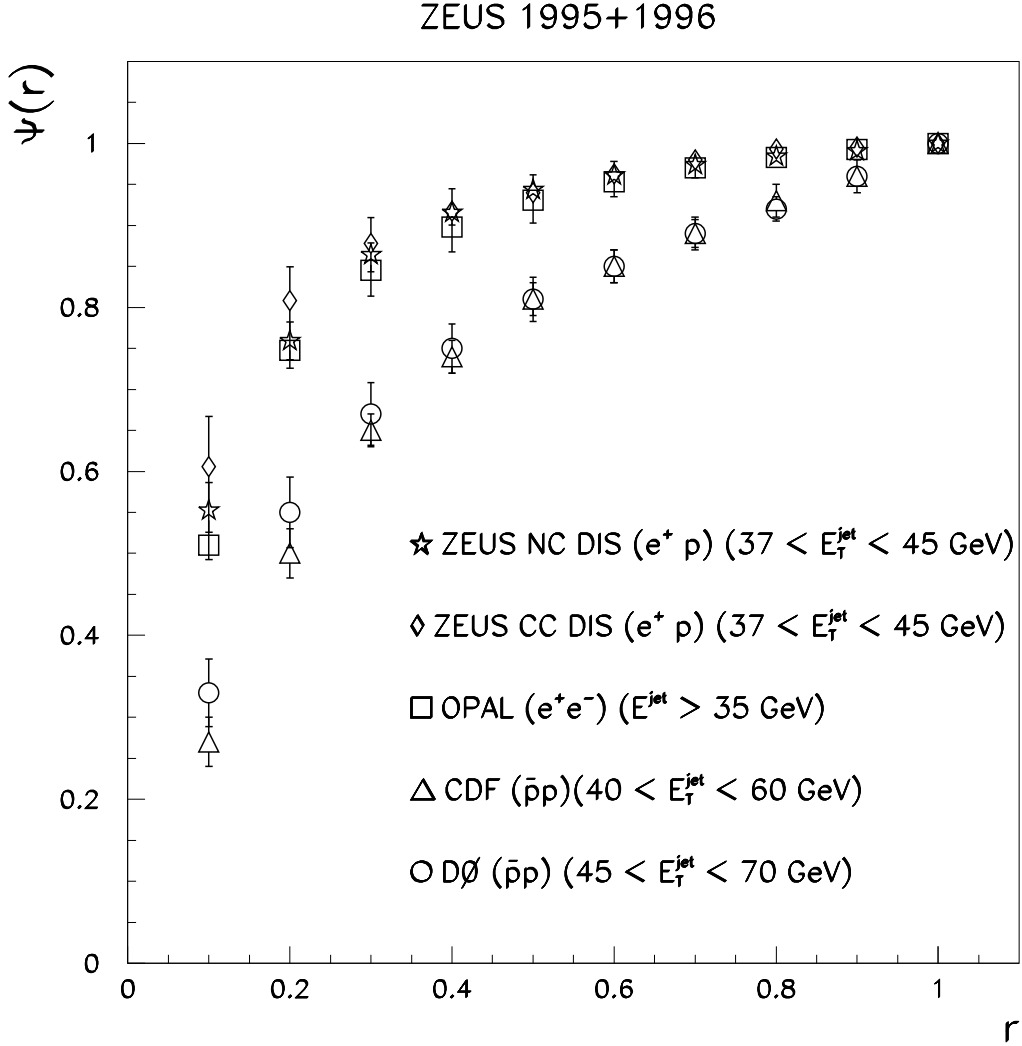,height=15cm}} 
\caption{Measured integrated jet shapes corrected to the hadron 
 level, $\psi(r)$, in neutral- (charged-) current DIS with $\q2 > 100$~\g2\ 
 for jets with $\etajet$ in the range between $-1$ and $2$ and
 $37< \etjet < 45$~GeV are shown as stars (diamonds). The measurements of
 jet shapes in $\bar{p}p$ collisions by CDF (triangles) and D\O\
 (circles) and in $e^+e^-$ interactions by OPAL (squares) are shown for
 comparison. 
\label{figdra9}}
\end{figure}

 The measured integrated jet shapes in DIS are compared to those in
$e^+e^-$ interactions and $\bar{p}p$ collisions in Figure~\ref{figdra9}. The
measured jets in DIS at HERA are found to be narrower than those in
$\bar{p}p$ collisions. The measurements in $\bar{p}p$ collisions have been
performed for jets with slightly higher energy than those in NC and CC DIS.
This difference cannot explain the discrepancy in the jet shapes
since the jets become narrower as the jet energy increases. As stated in
\cite{opal1}, most of the difference between the jet shapes in $e^+e^-$
interactions and $\bar{p}p$ collisions can be ascribed to the larger fraction of
gluon jets in the latter reaction. The comparison between the measured jet
shapes in DIS and $\bar{p}p$ collisions suggests that, also in this case, the
difference can be attributed to differences between quark and gluon jet
properties.

\section{Summary and conclusions}

 Measurements have been presented of the differential and integrated jet
shapes in neutral- and charged-current deep inelastic $e^+p$ scattering at
$\sqrt{s}=300$~GeV using data collected by ZEUS in 1995 and 1996. The 
jet shapes refer to jets at the hadron level with a cone radius of one 
unit in the $\etaphi$ plane and are given for the kinematic region
$Q^2 > 100$~GeV$^2$. Jets with $\etjet > 14$~GeV and $\etar$ have been 
considered. The jets become narrower as $\etjet$ increases. No 
significant $\etajet$ dependence of the jet shape has been observed. 
The measured jet shapes in neutral- and charged-current DIS are found to
be very similar. 

 The measurements of jet shapes have been compared to the predictions of
Monte Carlo generators using different models for the QCD radiation. The
colour-dipole model as implemented in {\sc ariadne} provides a reasonable
description of the measured jet shapes in all $\etajet$ and $\etjet$ regions
studied. The parton-shower approach without first-order QCD matrix-elements
predict jets which are slightly narrower at low $\etjet$ than those in the
data for all the $\etajet$ regions studied. The inclusion of first-order QCD
matrix-elements improves the description of the data for $\etajet >1$, but
leads to jets which are slightly broader for $\etajet<1$.

 The jets in neutral-current DIS are narrower than those in dijet
photoproduction but closer to those in direct-photon processes for the
same ranges in jet transverse energy and pseudorapidity. The jets in DIS
are found to be narrower than those in $\bar{p}p$ collisions. This
difference can be attributed to a larger contribution of gluon jets in
$\bar{p}p$ collisions. The measured jet shapes in neutral- and
charged-current DIS are similar to those in $e^+e^-$ interactions for
comparable ranges of jet transverse energy. Since the jets in $e^+e^-$
interactions and deep inelastic $e^+p$ scattering are predominantly quark
initiated, the similarity in the jet shapes indicates that the pattern of
QCD radiation within a quark jet is to a large extent independent
of the hard scattering process in these reactions.

\vspace{0.5cm}
\noindent {\Large\bf Acknowledgements}
\vspace{0.3cm}

 The strong support and encouragement of the DESY Directorate have been 
invaluable. The experiment was made possible by the inventiveness and the 
diligent efforts of the HERA machine group.  The design, construction and 
installation of the ZEUS detector have been made possible by the
ingenuity and dedicated efforts of many people from inside DESY and
from the home institutes who are not listed as authors. Their 
contributions are acknowledged with great appreciation. 



\end{document}